# Limited Perturbation of a DPPC Bilayer by Fluorescent Lipid Probes: A Molecular Dynamics Study

*Frederick A. Heberle [a]\*, David G. Ackerman [b], and Gerald.W. Feigenson [b]*

[a] Neutron Scattering Science Division, Oak Ridge National Laboratory, Oak Ridge, TN 37831, USA

[b] Field of Biophysics, Cornell University, Ithaca, NY 14853, USA

\*To whom correspondence should be addressed. E-mail: heberlefa@ornl.gov, Phone: (865) 576-8802, Fax: (865) .

# ABSTRACT

The presence and the properties of lipid bilayer nanometer-scale domains might be important for understanding the membranes of living cells. We used molecular dynamics (MD) simulations to investigate perturbations of a small patch of fluid-phase DPPC bilayer upon incorporation of fluorescent indocarbocyanine lipid probes commonly used to study membranes (DiI-C12:0, DiI-C18:0, or DiI-C18:2). In simulations containing 1 probe per 64 total lipids in each leaflet, an 8 - 12% decrease in chain order is observed for DPPC molecules in the solvation shell closest to the probe, relative to a pure DPPC bilayer. A ~5% increase in chain order is seen in the next three shells, resulting in a small overall increase in average DPPC chain order. In simulations with 1 probe per 256 total lipids in each leaflet, average DPPC chain order is unaffected by the probe. Thus, these DiI probes cause an oscillatory perturbation of their local environment but do not strongly influence the average properties of even "nanoscopic" lipid phase domains.

**KEYWORDS**:

# INTRODUCTION

Probe-based studies greatly aid our understanding of lipid membranes, and fluorescent probes in particular have proven useful. Bilayer properties studied with fluorescence techniques include order (1), hydration and polarity (2), electrostatic potential (3), lipid lateral diffusion (4), and phase state (5). In recent years, fluorescence microscopy has played an important role in elucidating the lateral organization of cell membranes. Fluorescence studies continue to drive the raft field, as the search for mechanisms responsible for the small size and transient nature of rafts has widened to include critical behavior (6) and stable nanoscopic phase domains (7).

Fluorescence probes report on their very local environment, and proper interpretation of fluorescence data requires understanding the extent to which the probe itself perturbs its local environment. Clearly, any such perturbations must be assessed independently from the information reported by the probe, and several techniques have been used for this purpose. Differential scanning calorimetry (8) and $^2$H NMR (9) have been used to detect changes in melting transition temperature in bilayers doped with fluorescent probes, X-ray diffraction has been employed to measure differences in average bilayer structure (10), and $^1$H NMR has been used to measure changes in motional freedom of the lipid (11). Each of these techniques reports on average properties of a large number of lipids. Typically, significant changes are not detected until the probe concentration exceeds several percent, although exceptions are reported (9). Bulk properties are inherently insensitive to the very local perturbations induced by a probe at the dilute concentrations (< 0.1 mol %) typically found in spectrophotometric experiments. Two different but related questions can be posed: To what extent does the probe report on a perturbed local environment? And, how far out from the probe is the lipid perturbed?

Molecular dynamics (MD) simulations can provide insight into bilayer structure and dynamics that might otherwise be impossible to obtain experimentally. All-atom simulations offer a unique way of characterizing spatial dependence of perturbations induced by a probe, including local changes. MD studies of bilayers containing fluorescent probes have appeared in the literature, aimed at understanding the location and dynamics of the fluorophore and its perturbative effects within the bilayer. Reviewing several of these studies, Loura and Ramalho emphasized the important distinction between first-shell lipids and the average properties of all lipids in the simulation (12). In a recent report, the indocarbocyanine probe DiI-C18:0 was found to increase the *average* order and thickness of a fluid DPPC bilayer (13). Here, we used MD to examine the spatial dependence of perturbation from three types of DiI that differ by alkyl chains being 12:0, 18:0, or 18:2. We focus on *local* changes in DPPC order. The three simulated bilayers at 50 °C were composed of 126 DPPC and 2 DiI (one per leaflet), with a fourth system composed of 128 DPPC for comparison. 100 ns simulations were performed, and the final 90 ns analyzed. We found that the behavior of first-shell lipids was highly variable across multiple simulations. However, by running 12 independent simulations for each system studied, we can report 95% confidence inverals on the resulting data sets that describe the probe-induced perturbation of the local lipid environment.

**SIMULATION METHODS**

*128-lipid simulations*

MD simulations were performed using version 4.5.1 of the Groningen Machine for Chemical Simulations (GROMACS) and the ffg53a6_lipid force field. The PDB file of a template bilayer consisting of 128 DPPC molecules was obtained from the Tieleman website (14). A DiI-C18:0 PDB file was constructed using the PRODRG2 server (15), and its charge distribution parameters were modified based on Gullapalli et al. (13). A DiI-C12:0 PDB was constructed from the DiI-C18:0 file by truncating the alkyl chains. Similarly, a DiI-C18:2 PDB file was constructed from the DiI-C18:0 file by replacing the single bonds between C9-C10 and C12-C13 in the alkyl chains with *cis* double bonds, using parameters from the oleoyl chain of POPC (14). Initial simulations were viewed in PyMOL to ensure planar rigidity of the DiI chromophore as well as the proper double bond characteristics of DiI-C18:2.

Simulations containing 128 total lipids (systems A-D, Table 1) were performed following Gullapalli et al. (13), with minor modifications. One DPPC per leaflet was removed from the template bilayer and replaced with a reference "probe" molecule: DiI-C18:0, DiI-C18:2, DiI-C12:0, or DPPC. The two probes were inserted in a position that maximized their separation distance. For DiI-C18:0 simulations (system B), probes were inserted with headgroups protruding slightly into the water. During equilibration, the headgroups were quickly drawn into the bilayer interior, confirming the observations of Gullapalli et al. (13). For systems A, C, and D, the probes were inserted such that their headgroups were at approximately the same level as the DPPC headgroups. A total of 12 simulations for each reference molecule were set up in this manner, differing only in the initial orientation of the inserted probe. The bilayers were expanded using INFLATEGRO (14,16), and energy minimized to reduce unfavorable interactions created during the probe insertion step. The system was then repeatedly shrunk via INFLATEGRO, and energy minimized. The sequence of shrinking and energy minimization was repeated until the average area per lipid was within the range of accepted values for DPPC (17). Solvation was then performed using the simple point charge (SPC) water model, and for the DiI insertions, two chloride ions were added to cancel the net charge of the system. This was again followed by an energy minimization. Next, NVT temperature equilibration was performed at 323 K ($L_d$ phase temperature for DPPC) for 10 ns using a Nosé-Hoover thermostat, followed by NPT pressure equilibration for 10 ns at 1 atm, using a Nosé-Hoover thermostat, Parrinello-Rahman barostat, and semiisotropic (zero surface tension) pressure coupling. Finally, 100 ns production runs were performed under NPT ensemble conditions, and the final 90 ns were analyzed.

Periodic boundary conditions were applied in all three spatial directions, with x and y in the plane of the bilayer and z normal to this bilayer plane. Bond lengths were constrained using the LINCS algorithm and the Particle-Mesh Ewald (PME) method was used for electrostatic interactions, with cubic interpolation order 4, a direct space cutoff of 1.2 nm, and a Fourier transform grid spacing of 0.12 nm. A short range interaction cutoff of 1.2 nm, and a Lennard-Jones interaction cutoff of 1.2 nm were also applied. The equations of motion were integrated via the leap-frog algorithm with a timestep of 2 fs.

A representative simulation snapshot from system B is shown in Fig. 1, and representative data from systems A-D are shown in Fig. 2.

*Data analysis*

Our primary objective is to understand the size and spatial extent of probe perturbations. We decided to examine average properties of the host lipid (fluid phase DPPC) as a function of distance from the probe molecule. Given this stated objective, there are many possible ways to analyze the MD trajectory data. Here, we have partitioned the simulation box into zones, each defined by a range of probe-lipid separation distances, and potentially different average properties. The zones can be thought of as "solvation shells" around the probe, each containing an integral number of DPPC lipids at any particular instant. If we define the $n^{\text{th}}$ solvation shell as an annulus containing $6n$ lipids *in an ensemble average of trajectory frames,* we need only to specify the inner and outer radii of each annulus that satisfies this condition. To simplify the comparison between systems, we have chosen to define a single, fixed set of annuli based on the pure DPPC system, as described below.

A simple way to partition the molecules into shells is to consider only the projections of the molecular centers-of-mass $\boldsymbol{\rho}$ onto the 2D plane of the bilayer (see Fig. 3). Taking the probe center-of-mass $\boldsymbol{\rho_P}$ as the origin of a 2D coordinate system, the probe-lipid separation distance $r$ is defined as follows:

$$r = |\boldsymbol{\rho_P} - \boldsymbol{\rho_L}| \qquad\qquad 1$$

where $\boldsymbol{\rho_L}$ is the lipid center-of-mass. Positional order is very short range in a fluid phase, and $\boldsymbol{\rho_L}$ are nearly randomly distributed. Under this assumption, shell $n$ has an average area of $6nA_L$, where $A_L$ is the average area of a host lipid. The annular radii are then defined as follows:

$$r_0 = \sqrt{A_L/\pi}$$

$$r_n = r_0 \left(1 + 6\sum_{i=1}^{n} i\right)^{1/2} = r_0\sqrt{1 + 3n(n+1)}$$

At any point in time, a lipid is considered to reside in shell $n$ if its center-of-mass falls within the inner and outer radii of the shell—that is, if $r_{n-1} \le r < r_n$ (with $r < r_0$ assigned to the first shell). This partitioning scheme is shown graphically in Fig. 3.

We emphasize that the shell radii are defined based only on *average* properties of DPPC: a single, fixed set of annuli is used for all frames and all systems studied. In addition to neglecting the short range positional order of a fluid, this definition of shells neglects the average area of the probe molecule (which will in general be different than that of the host lipid), as well as any distance-dependent perturbations of lipid areas. Though somewhat crude, the method nevertheless partitions the simulation box so that each shell contains an average of ~ $6n$ lipids, as shown in Table 2. Other partitioning schemes with fixed and consistent definitions for the shell radii are expected to give similar results. We note however that defining solvation shells on a per-frame basis using a Voronoi tessellation of lipid centers-of-mass results in a particularly interesting artifact: lipids with smaller areas (and greater order) are preferentially "squeezed" into higher shell numbers, relative to a partitioning scheme based strictly on distance from the probe. For this reason, the Voronoi tessellation should not be used to examine distance-dependence of probe perturbations.

Segmental order profiles for the *probe chains* are shown in Fig. 4. Segmental order profiles for DPPC chains in the first three solvation shells are shown in Figs. 5-7 for DiI-C18:0, DiI-C12:0, and DiI-C18:2, respectively.

### 512-lipid simulations

The minimum probe concentration that can be obtained in a symmetric bilayer with 128 total lipids is 1/64, or 1.6 mol %. It is generally recognized that such high probe concentrations are not desirable for studies of membrane phase behavior, and many spectroscopic studies use overall probe concentrations 10 or even 100 times more dilute than is used in our simulations of systems A-C. Nonetheless, if the actual organization of a lipid bilayer involves domains on the size scale of ~10 nanometers and thus ~100 lipids, a locally high probe concentration does occur for that nanodomain, however dilute the overall probe concentration might be. In order to examine the influence of simulation size and probe concentration, additional simulations were performed with 512 total lipids (systems E and F, Table 1). Bilayers were generated with a 2x2 tiling of the original 128 lipid template. Systems E and F were set up as previously described: probes were inserted to maximize their separation distance, with each probe's headgroup at the same height as the DPPC headgroups. The total production runtime for these larger systems was 50 ns; all other simulation details were identical to systems A-D.

**RESULTS AND DISCUSSION**

Fig. 8 shows mass density profiles of DPPC and DiI-C18:0 averaged over both leaflets. Probe structures (shown in Figs. 9 and 10) show a flat, extended DiI chromophore attached to two alkyl chains at the indole nitrogens. DiI are almost entirely hydrocarbon, but a positively charged chromophore confers weak amphiphilic character: they incorporate into the membrane in a preferred orientation, with chromophore toward the interface, and chains inserted among those of the host lipid. The headgroup is within the lipid hydrocarbon, in agreement with previous simulations (13,18). Mass density of selected probe and DPPC groups is shown in Fig. 11. Chromophore density peaks ~ 11 Å from the bilayer center, nearly independent of the probe alkyl chains. There is substantial interdigitation of DiI chains: probe 18:0 and 18:2 chains show density nearly 10 Å into the opposite leaflet, and extend 3-4 Å further on average than DPPC chains. In contrast, 12:0 chains exhibit similar interdigitation to DPPC chains.

We calculated carbon-deuterium order parameters for lipid chains as follows:

$$S_{CD} = \langle 3\cos^2\theta - 1 \rangle / 2 \qquad 3$$

where $\theta$ is the angle between the C-D bond and the magnetic field, chosen to be the membrane normal. The average segmental $S_{CD}$ profiles *for the probe chains* are shown in Fig. 4. All DiI chains are more disordered than DPPC chains. The saturated 12:0 and 18:0 chains have similar profiles, whereas the double bonds introduce significant disorder in the 18:2 chains. Both the location of the bulky chromophore among the DPPC chains and the substantial disorder of the probe chains raise the possibility of local disordering of DPPC chains.

We examined the spatial dependence of probe perturbations by dividing the bilayer into shells of lipid around the probe, and calculating average properties for DPPC in each shell. Briefly, shell $n$ is defined to contain on average $6n$ DPPC. The following discussion focuses on $S_{CD}$. Fig. 5 shows $S_{CD}$ profiles of DPPC in each shell surrounding DiI-C18:0. Order is decreased for all carbon segments in DPPC chains located closest to the probe. The small number of first-shell lipids results in high variability even in a 100 ns simulation, and in a few individual cases first-shell order is increased relative to pure DPPC. However, the average of multiple independent simulations shows that $S_{CD}$ for nearest-neighbor lipids is ~ 8% smaller than unperturbed DPPC, and that the result is statistically significant. A similar effect is observed for 12:0 and 18:2 chains (Figs. 6 and 7), though the magnitude of the decrease is 12% for 18:2 chains (Table 3). Evidently the DiI chromophore, whether chains are 12:0, 18:0, or 18:2 facilitates greater motional freedom of DPPC chains, but only those in closest proximity. Interestingly, increased order is seen in the second and third shells, with $S_{CD}$ values 3-7% greater than unperturbed DPPC. Averaging over all DPPC in the bilayer irrespective of shell obscures the distinction between the first-shell and more distant lipids—$S_{CD}$ shows a small but significant increase for each probe studied when distance depencence is not taken into account, consistent with published MD results for DiI-C18:0 influence averaged over all lipids (13,18).

The *qualitative* difference between the local and overall average probe perturbation is a suprising result, and suggests a nonmonotonic decay to the unperturbed state. Preliminary results for a single DiI-C18:0 probe in a larger leaflet (1 probe per 256 total lipids) lend support to a gently oscillating perturbation profile (Table 4), though better statistics are needed. Oscillations in bilayer thickness near protein or lipid inclusions are predicted theoretically (19-21), and arise from an interplay between the perturbing molecule and the mechanical properties of the unperturbed bilayer. Importantly, we find that for these larger bilayers at lower probe concentration, the disordering effect is still ~ −10% for nearest-neighbor DPPC, but the overall average DPPC order is nearly identical to that of a pure DPPC bilayer. A reasonable conclusion is that overall properties of a small domain containing as few as ~ 100-300 lipids are not significantly perturbed by a single DiI probe.

Representative data from systems E and F are shown in Fig. 12. We find that upon doubling the box size, average properties (e.g., area per lipid and mass distributions of lipid and probe) are unchanged to within error. Table 4 shows average $S_{CD}$ for DPPC acyl chains in shells around DiI-C18:0, expressed as a fraction of the average value in a pure DPPC bilayer. Values for the first four shells (which extend nearly to the edges of the smaller simulation box) are similar for the smaller and larger systems. Importantly, the average order parameter decays to the unperturbed, pure DPPC values in system E, and the overall average bilayer order is essentially the same as in pure DPPC. However, given the poorer statistics of the larger simulations, we cannot determine the perturbation length with any certainty.

**CONCLUSIONS**

MD simulations are a powerful tool for examining the spatial extent of bilayer perturbations of a fluorescent lipid analog. We find that multiple independent simulations are required to faithfully report on the small number of lipids in the first shell around the probe. Furthermore, when analysis is limited to only average bilayer properties, local probe effects can be misunderstood. This problem is clearly demonstrated by our observation of an overall increase in bilayer order in the presence of DiI, despite the probe's local disordering of DPPC chains. This result suggests an oscillating perturbation profile, but caution is warranted in this interpretation: the periodic systems in this study represent a highly correlated superlattice of probe molecules at concentrations found in nanometer-scale domains, but higher than are typical of the overall mixture. We are currently investigating larger systems and lower probe concentrations to help understand this interesting result.

**ACKNOWLEDGMENT**

Support was from research awards from the NIH R01 GM077198 and the NSF MCB 0842839 (to G.W.F.). We thank Alan Grossfield, Jonathan Amazon, Juyang Huang, and Hari Muddana for helpful discussions.

**TABLE 1** Selected simulation parameters for the systems studied.

| System | # DPPC | Inserted Molecules | Probe mole fraction | # Water | # Cl⁻ | Production Runtime (ns) | # Simulations |
|--------|--------|--------------------|--------------------|---------|-------|------------------------|---------------|
| A | 126 | DiI-12:0 (2) | 1/64 | ~2530 | 2 | 100 | 12 |
| B | 126 | DiI-18:0 (2) | 1/64 | ~2450 | 2 | 100 | 12 |
| C | 126 | DiI-18:2 (2) | 1/64 | ~2450 | 2 | 100 | 12 |
| D | 126 | DPPC (2) | 0 | ~2530 | 0 | 100 | 12 |
| E | 510 | DiI-18:0 (2) | 1/256 | ~10140 | 2 | 50 | 3 |
| F | 510 | DPPC (2) | 0 | ~10140 | 0 | 50 | 4 |

**TABLE 2** Solvation shell definitions calculated from Equation 2, using an average DPPC area of 61.3 Å$^2$ (calculated from pure DPPC simulations). Also shown are the average number of lipids assigned to each shell for the systems studied.

| Shell | Inner radius (Å) | Outer radius (Å) | Average number of host lipids in shell | | | |
|-------|-----------------|-----------------|-----------|-----------|-----------|-----------|
| | | | System A | System B | System C | System D |
| 1 | 0 | 11.7 | 6.0 | 6.0 | 6.0 | 5.9 |
| 2 | 11.7 | 19.3 | 12.3 | 12.1 | 12.0 | 12.1 |
| 3 | 19.3 | 26.9 | 18.2 | 18.2 | 18.3 | 18.0 |

**TABLE 3** Average S$_{CD}$ for DPPC acyl chains as a function of distance from DiI, expressed as a fraction of the average value in a pure DPPC bilayer (95% confidence in parentheses).

| System | Shell 1 | Shell 2 | Shell 3 | Average |
|--------|---------|---------|---------|---------|
| DiI-C12:0 | 0.92 (0.02) | 1.04 (0.02) | 1.06 (0.01) | 1.04 (0.01) |
| DiI-C18:0 | 0.91 (0.03) | 1.03 (0.02) | 1.06 (0.01) | 1.04 (0.01) |
| DiI-C18:2 | 0.88 (0.03) | 1.03 (0.02) | 1.07 (0.01) | 1.04 (0.01) |

**TABLE 4** Average $S_{CD}$ for DPPC acyl chains in 128- and 512-lipid systems as a function of distance from DiI-C18:0. Order is expressed as a fraction of the average value in a pure DPPC bilayer (95% confidence intervals in parentheses).

| System | # DPPC | Shell 1 | Shell 2 | Shell 3 | Shell 4 | Shell 5 | Shell 6 |
|--------|--------|---------|---------|---------|---------|---------|---------|
| B | 126 | 0.91 (0.03) | 1.03 (0.02) | 1.06 (0.01) | 1.06 (0.02) | -- | -- |
| E | 510 | 0.88 (0.08) | 1.00 (0.06) | 1.02 (0.04) | 1.04 (0.05) | 1.03 (0.04) | 1.02 (0.05) |

| System | Shell 7 | Shell 8 | Shell 9 | Shell 10 | Shell 11 | Shell 12 | Bilayer avg |
|--------|---------|---------|---------|----------|----------|----------|-------------|
| B | -- | -- | -- | -- | -- | -- | 1.04 (1) |
| E | 1.02 (0.04) | 1.00 (0.03) | 0.96 (0.05) | 0.98 (0.07) | 0.96 (0.06) | 0.96 (0.18) | 1.01 (1) |

**FIGURE CAPTIONS**

**Figure 1**.       Representative simulation snapshot reveals DiI-C18:0 location in the bilayer. DPPC headgroups are shown in space-filling representation, with the rest of the molecule shown in stick representation. The DiI chromophore (green) is found beneath the DPPC headgroups, predominantly within the hydrophobic interior of the bilayer. Terminal methyls (purple) reveal interdigitation of DiI chains into the opposite leaflet. Chloride ions (cyan) were added to neutralize the positive charge of the DiI chromophore.

**Figure 2**.       Representative data from systems A-D. Top row, area/lipid during a representative 100 ns simulation. Average values over the final 90 ns (red lines) are consistent with literature values for DPPC (17).  Bottom row, transverse bilayer position with respect to bilayer midplane (z=0, red line) of DPPC nitrogen (blue) and phosphate (yellow), and center-of-mass of the DiI chromophore (green) and entire DiI molecule (black).

**Figure 3**.       Partitioning of DPPC into solvation shells. Probe and lipid centers-of-mass (circles) are projected onto the plane of the bilayer. The probe center-of-mass (red) is taken to be the origin, and shell inner and outer radii $\{r_i\}$ are defined such that shell $n$ contains on average $6n$ lipids (see Equation 2). A host lipid is considered to be contained in shell $n$, if $r_{n-1} \leq r < r_n$ (and in the first shell, if $r < r_0$).

**Figure 4**.       Segmental order parameter profiles for DiI probes. DiI-C18:0 (blue short dashed) and DiI-C12:0 (red dotted) show similar order. Double bonds at the 9-10 and 12-13 position result in significant disordering of the DiI-C18:2 chains (green long dashed). DPPC *sn*-1 (black solid) and *sn*-2 (black dot-dashed) profiles are shown for reference.

**Figure 5**. Order parameter profiles for DPPC as a function of distance from DiI-C18:0. Average unperturbed DPPC *sn*-1 (A) and *sn*-2 (B) profiles (solid black), compared to chains in first (dotted red), second (short dash green), and third (long dash blue) shells around DiI-C18:0. Also shown are the highly variable first-shell DPPC profiles around individual probes (solid gray). The uppermost profile is a statistical outlier and was omitted from calculations of average order.

**Figure 6**. DPPC order parameter profiles versus distance from DiI-C12:0. Average unperturbed DPPC *sn*-1 (A) and *sn*-2 (B) profiles (solid black), compared to chains in first (dotted red), second (short dash green), and third (long dash blue) shells around DiI-C12:0. DPPC profiles around individual probes (solid gray) are shown to demonstrate the high variability.

**Figure 7**. DPPC order parameter profiles versus distance from DiI-C18:2. Average unperturbed DPPC *sn*-1 (A) and *sn*-2 (B) profiles (solid black), compared to chains in first (dotted red), second (short dash green), and third (long dash blue) shells around DiI-C18:2. DPPC profiles around individual probes (solid gray) are shown to demonstrate the high variability.

**Figure 8.** Average single-leaflet mass density profiles for DPPC (A) and DiI-C18:0 (B). Bilayer midplane at z=0.

**Figure 9**. Chemical structures of lipids used in this study.

**Figure 10**.    Color coding of molecular structures used in Figs. 8 and 11. DPPC: nitrogen (blue), phosphorous (yellow), glycerol carbon (green), carbonyl carbon (red), methylene carbon (black), terminal methyl (purple). DiI: chromophore (green), headgroup methyl (red), nitrogen (blue), methylene carbon (black), alkene carbon (orange), terminal methyl (purple).

**Figure 11**.    Mass density profiles reveal DiI location among DPPC chains. DiI chromophore resides beneath the DPPC carbonyl independent of probe chain structure (12:0 dotted, 18:0 dashed, 18:2 dot-dash). Probe terminal methyl distribution shows significant interdigitation for longer chain species. DPPC density scaled 20x for clarity. Bilayer midplane at z=0.

**Figure 12**.    Representative data from systems E and F. Top row, area/lipid during a representative 50 ns simulation. Average values over the final 40 ns (red lines) are consistent with literature values for DPPC (4) and similar to values from 128-lipid simulations (systems A-D, see Fig. 2). Bottom row, transverse bilayer position with respect to bilayer midplane (z=0, red line) of DPPC nitrogen (blue) and phosphate (yellow), and center-of-mass of the DiI chromophore (green) and entire DiI molecule (black).

**Figure 1**

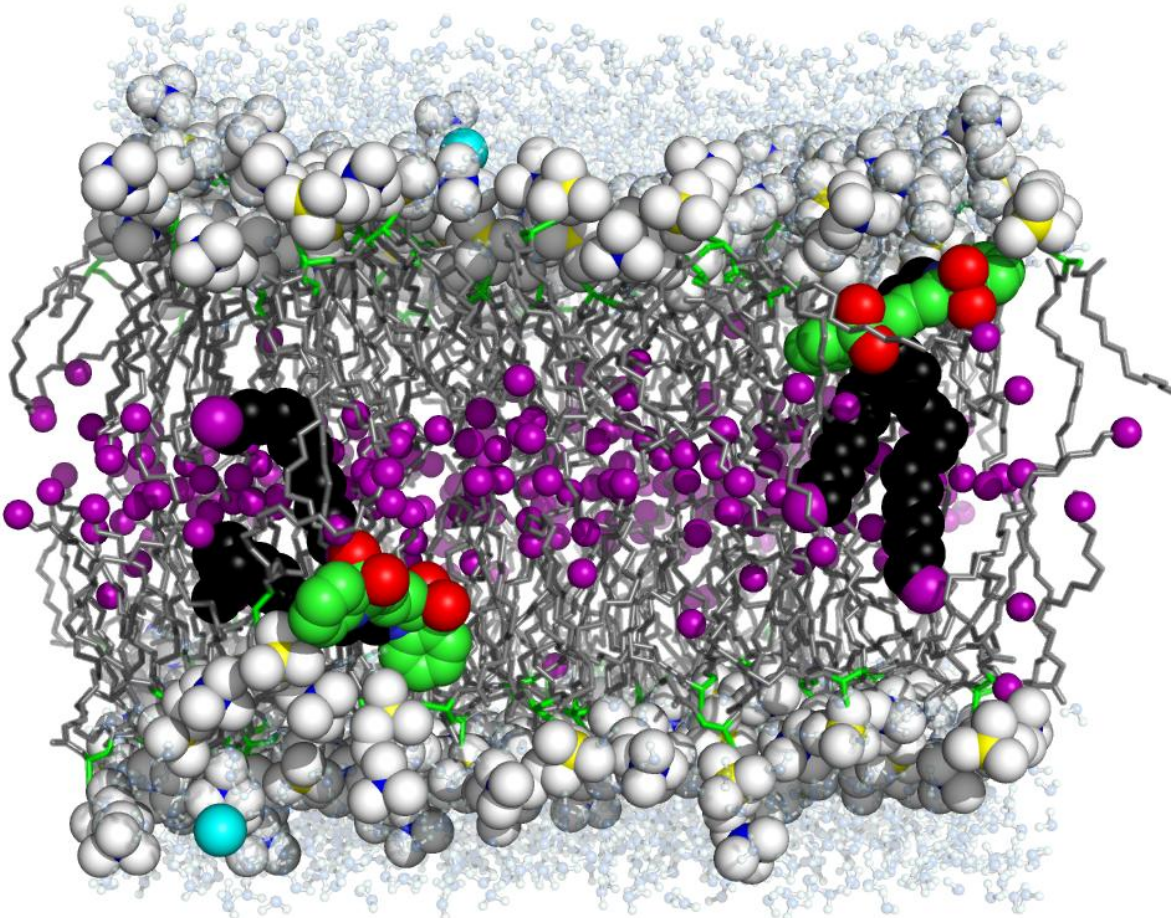

**Figure 2**

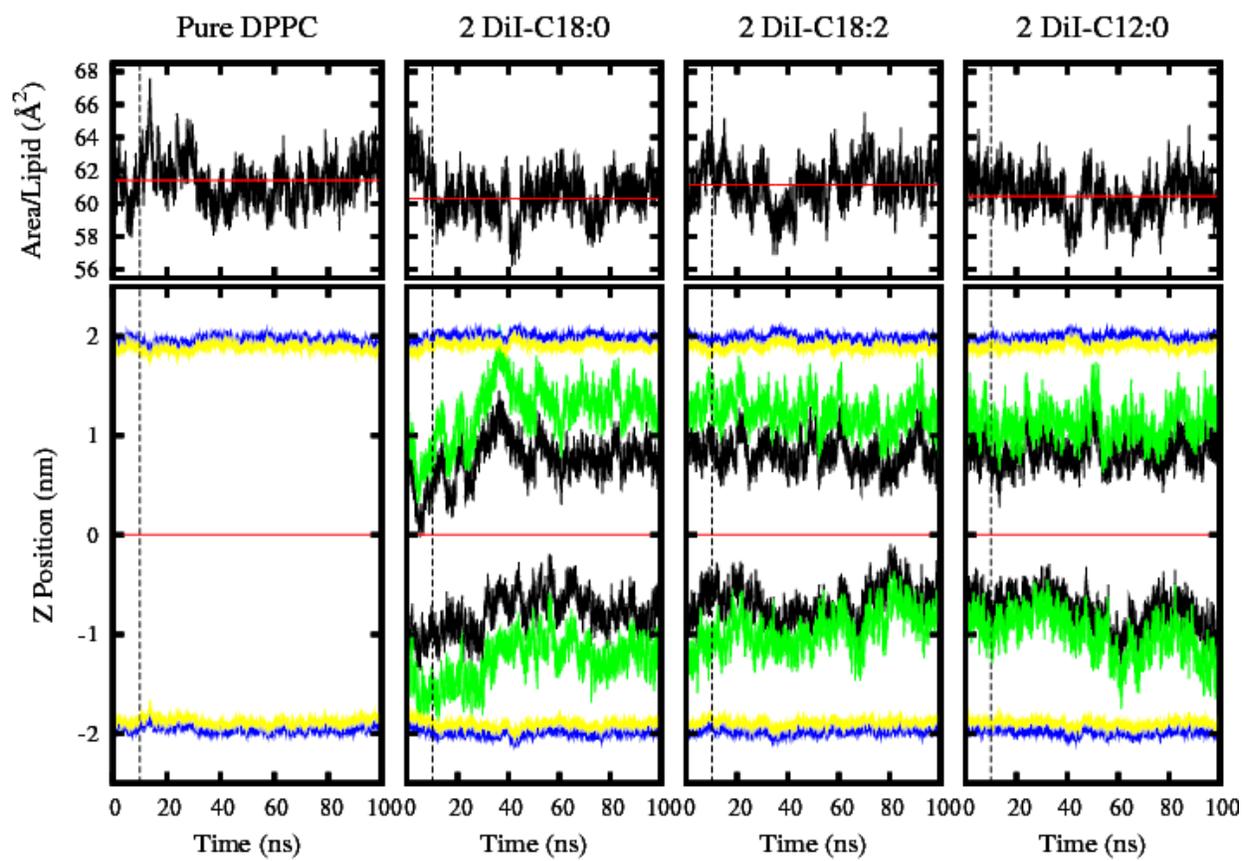

**Figure 3**

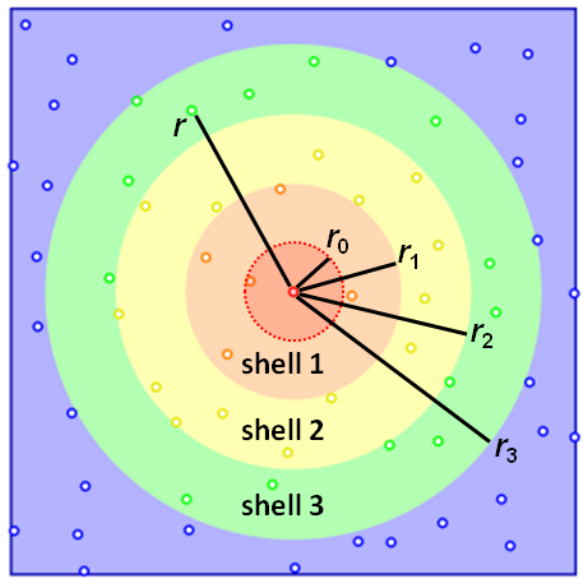

**Figure 4**

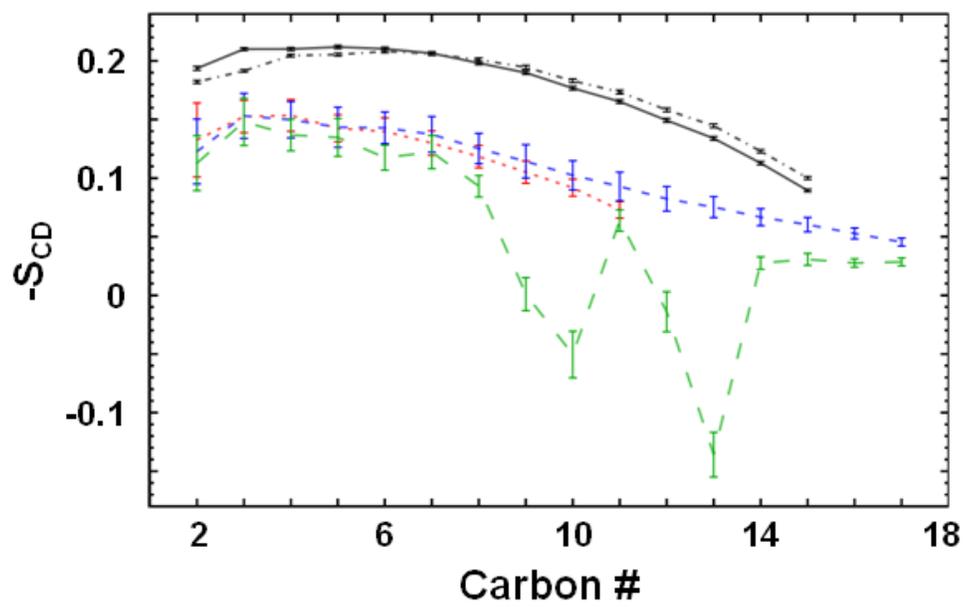

**Figure 5**

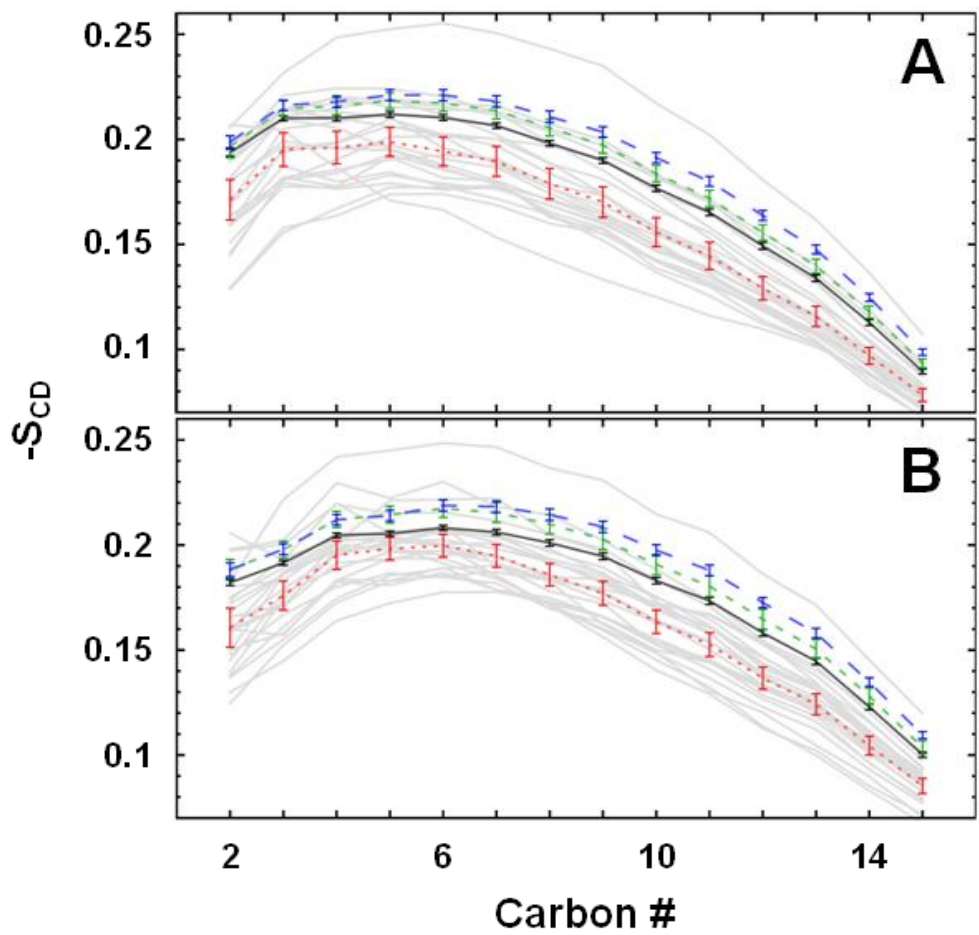

**Figure 6**

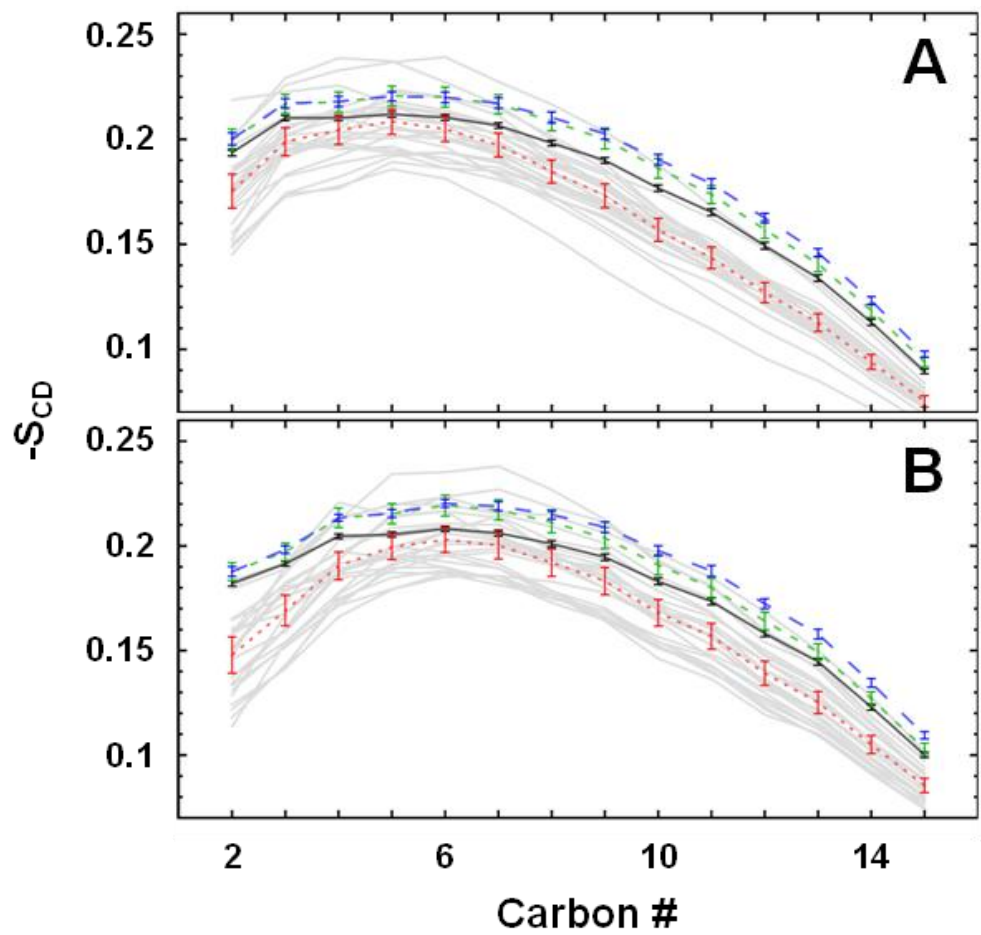

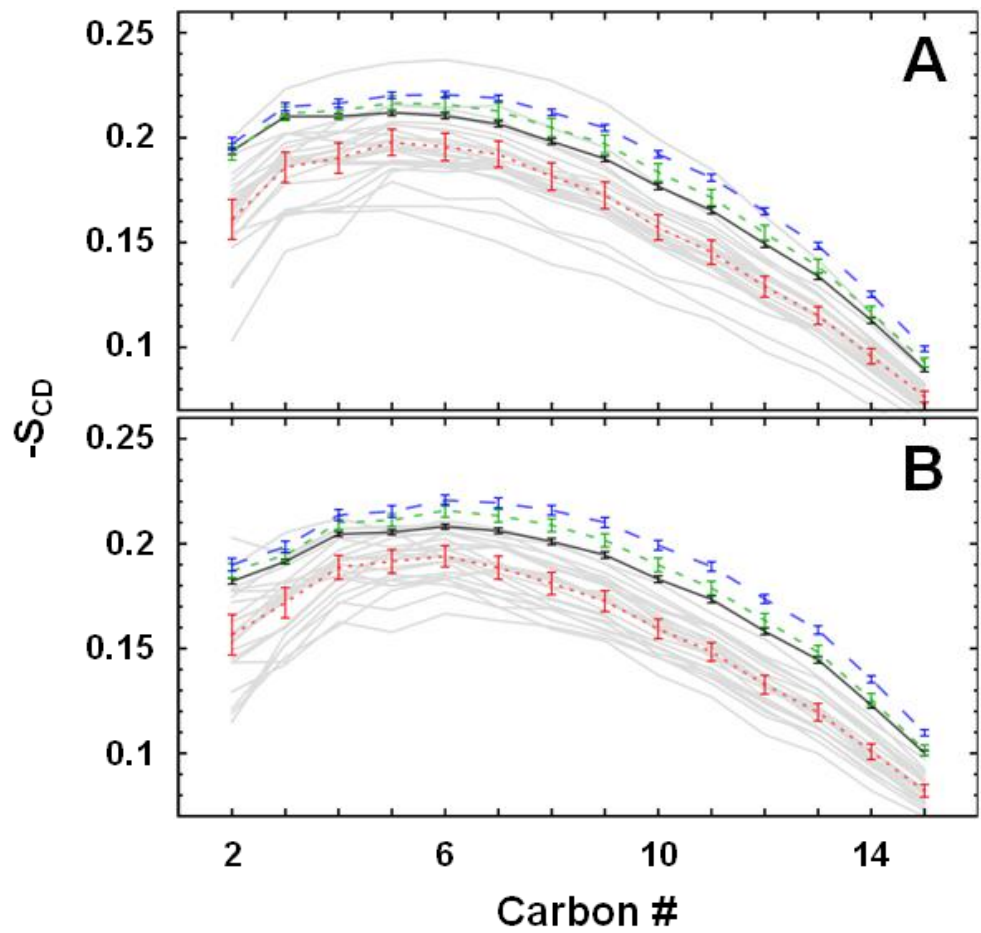

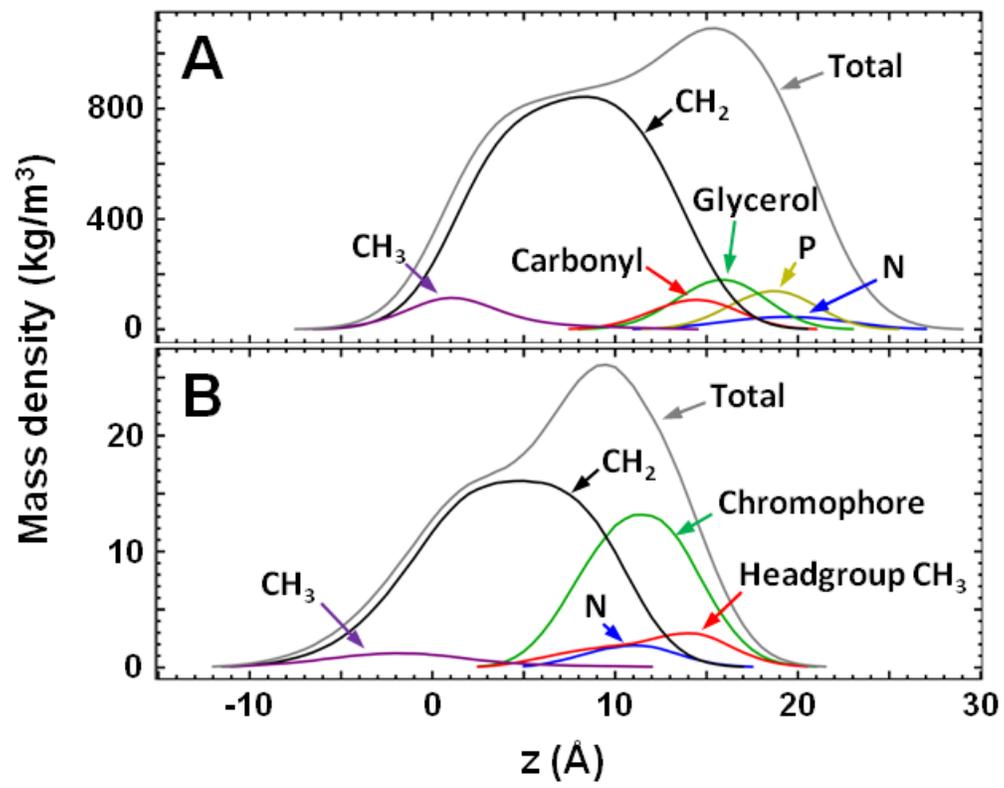

**DPPC**

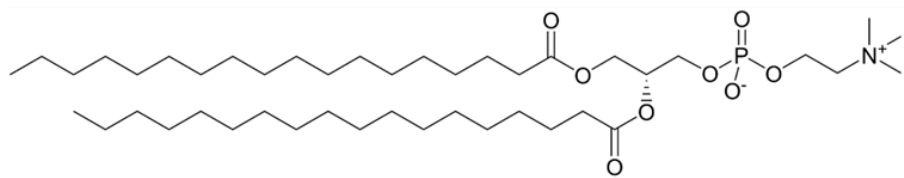

**DiI Chains**

**DiI Chromophore**

**C18:0** 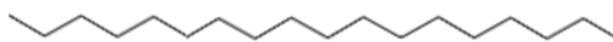

**C18:2** 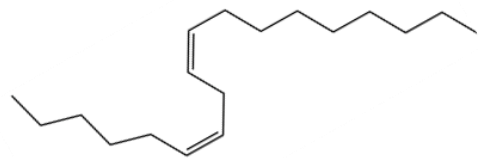

**C12:0** 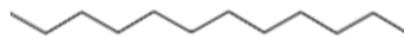

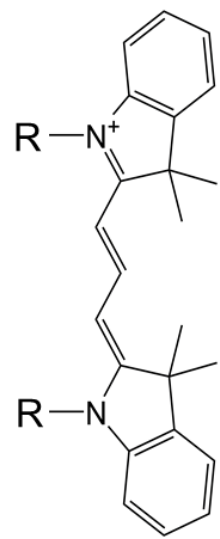

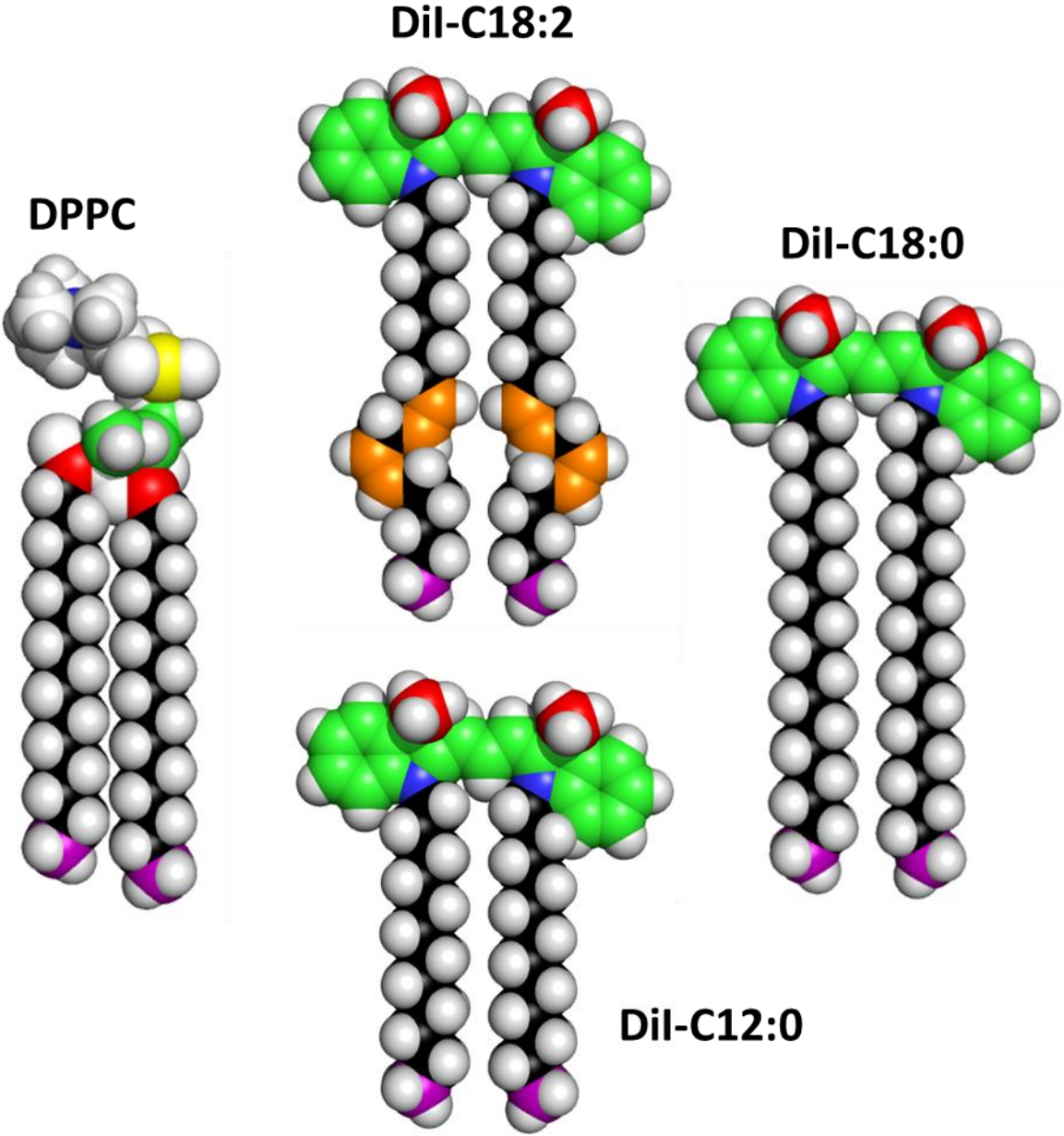

**DPPC**

**DiI-C18:2**

**DiI-C18:0**

**DiI-C12:0**

**Figure 11**

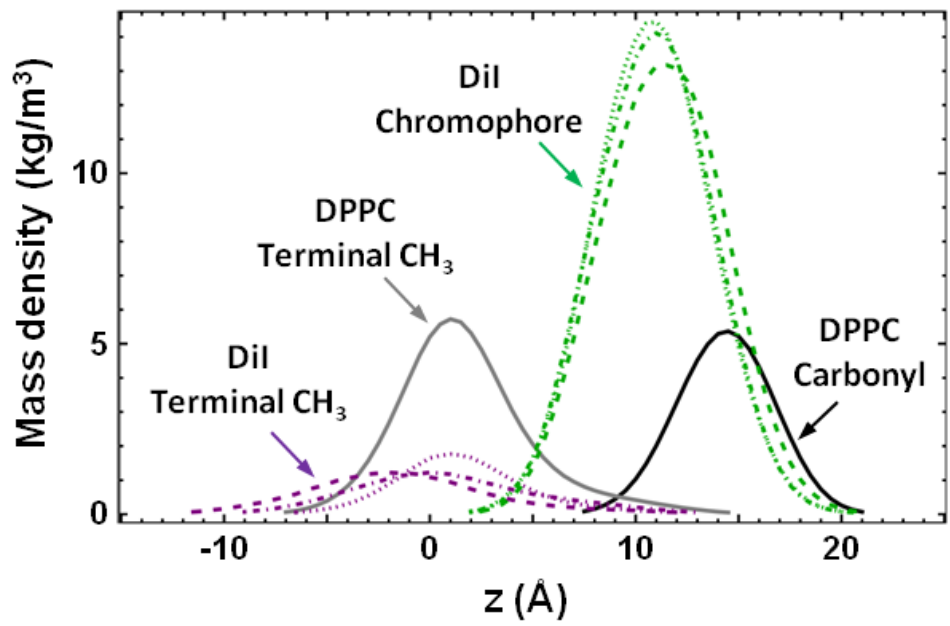

**Figure 12**

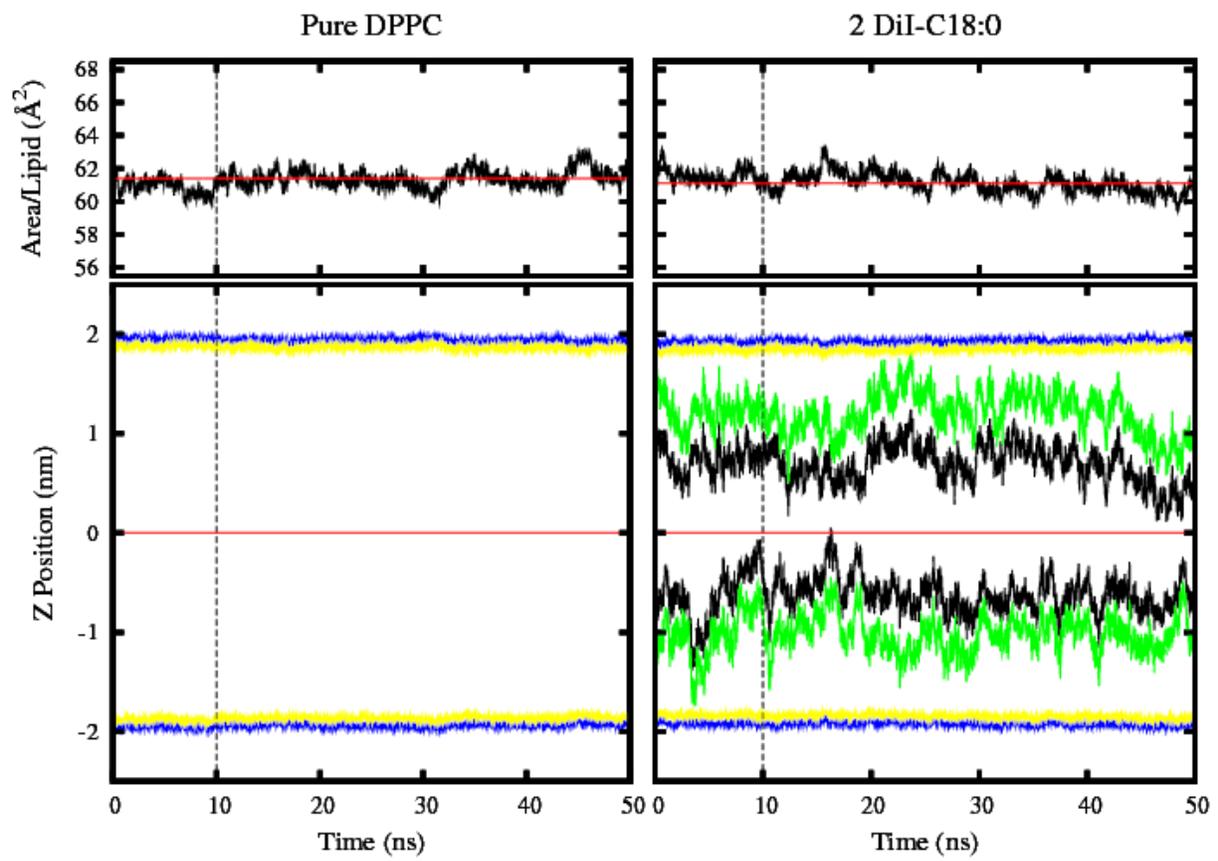